\documentclass[12pt, a4paper, twoside]{article}

\pdfoutput=1

\usepackage[tracking=true,expansion=true,stretch=15,shrink=15]{microtype}
\DeclareMicrotypeSet*[tracking]{my}
{ font = */*/*/sc/* }
\SetTracking{encoding = *, shape = sc }{ 45 }
\SetProtrusion{encoding={*},family={bch},series={*},size={6,7}}
{1={ ,750},2={ ,500},3={ ,500},4={ ,500},5={ ,500},
	6={ ,500},7={ ,600},8={ ,500},9={ ,500},0={ ,500}}
\SetExtraKerning[unit=space]
{encoding={*}, family={qhv}, series={b}, size={large,Large}}
{1={-200,-200},
	\textendash={400,400}}

\usepackage{setspace}
\usepackage[
	left=1.3in,
	right=1.3in,
	top=0.7in,
	bottom=0.7in,
	includeheadfoot
]{geometry}
\usepackage[disable]{todonotes}
\usepackage{lineno}
\usepackage[T1]{fontenc}
\usepackage{amssymb}
\usepackage{booktabs}
\usepackage{amsmath}
\usepackage{amsthm}
\usepackage{dsfont}
\usepackage{bm}
\usepackage{enumerate}
\usepackage{graphicx}
\usepackage{subfigure}
\usepackage{float}
\usepackage{tikz}
\usepackage{dcolumn}
\usetikzlibrary{shapes,arrows,decorations.pathmorphing,graphs,positioning,backgrounds}
\usepackage[round,comma]{natbib}
\usepackage{hyperref}
\usepackage{nameref}

\hypersetup{
	colorlinks   = true, %Colours links instead of ugly boxes
	urlcolor     = blue, %Colour for external hyperlinks
	linkcolor    = blue, %Colour of internal links
	citecolor   = blue %Colour of citations
}

\allowdisplaybreaks
\begin{document}

\title{Probabilistic Temperature Forecasting with a Heteroscedastic Autoregressive Ensemble Postprocessing model}

\author{A. M\"oller\footnote{Corresponding author, Institute of Applied Stochastics and Operations Research, Clausthal University of Technology, Clausthal-Zellerfeld, Germany, (email:\href{mailto:annette.moeller@tu-clausthal.de}{annette.moeller@tu-clausthal.de})},
J\"{u}rgen Gro{\ss}\footnote{Institute for Mathematics and Applied Informatics, University of Hildesheim, Germany}}

\date{\hspace{3pt} \normalsize\today}

\maketitle

%---------------------------------------------------------%

\begin{abstract}
Weather prediction today is performed with numerical
weather prediction (NWP) models. These are deterministic simulation models
describing the dynamics of the atmosphere, and evolving the current conditions
forward in time to obtain a prediction for future atmospheric states. To
account for uncertainty in NWP models it has become common
practice to employ ensembles of NWP forecasts. However, NWP
ensembles often exhibit forecast biases and dispersion errors, thus require
statistical postprocessing to improve reliability of the ensemble forecasts.
This work proposes an extension of a recently developed postprocessing model
utilizing autoregressive information present in the forecast error of the raw
ensemble members. The original approach is modified to let the variance parameter depend on the ensemble spread, yielding a two-fold heteroscedastic model. Furthermore, an additional high-resolution forecast is included into the postprocessing model, yielding improved predictive performance. Finally, it is outlined how the autoregressive model can be utilized to postprocess ensemble forecasts with higher forecast horizons, without the necessity of making fundamental changes to the original model.
We accompany the new methodology by an implementation within the \texttt{R} package \texttt{ensAR}
to make our method available for other researchers working in this area. To
illustrate the performance of the heteroscedastic extension of the autoregressive model, and its use for higher forecast horizons we present a case study for a data set containing 12 years of temperature forecasts and observations over Germany. The case study indicates that the autoregressive model yields particularly strong improvements for forecast horizons beyond 24 hours.\\[12pt]
\end{abstract}

%---------------------------------------------------------%

%\pagestyle{scrheadings}
%\clearscrheadings
%\lohead{Heteroscedastic Autoregressive Temperature Forecasting}
%\rohead{\pagemark}
%\lehead{M\"oller, Gro{\ss}}
%\rehead{\pagemark}

\section{Introduction}\label{sec:intro}

Today, weather prediction is based on numerical weather prediction (NWP) models. Such models are deterministic in nature and represent the dynamical physics of the atmosphere by a set of differential equations. The current state of the atmosphere is evolved forward in time to predict future atmospheric states. The solutions strongly depend on the initial conditions and model formulations. Thus, NWP models suffer from several sources of uncertainties.
Common practice in addressing these uncertainties is the use of ensemble prediction systems (EPS). The NWP model is run multiple times, each time with variations in the model parameterizations and/or initial and boundary conditions \citep{GneitingRaftery2005, LeutbecherPalmer2008}.
A forecast ensemble can be viewed as probabilistic forecast that allows to assess forecast uncertainty \citep{Palmer2002}. However, in practise, NWP ensembles exhibit forecast biases and dispersion errors, and require statistical postprocessing to improve calibration and forecast skill by utilizing recently observed forecast errors and observations. An additional benefit is that statistical postprocessing can yield full predictive probability distributions \citep{GneitingRaftery2005, gneiting2014probabilistic, WilksHamill2007}.

Statistical postprocessing models have enjoyed increasing popularity and success during the last decades and a variety of models tailored to specific problems have been developed. Many of the recently proposed models are extensions and modifications of two generic state-of-the-art models, namely the Ensemble Model Output Statistics approach (EMOS; \citealp{gneiting2005calibrated}) and the Bayesian Model Averaging (BMA; \citealp{Raftery&2005}).
The original EMOS and BMA models were designed for Gaussian distributed weather quantities, and a variety of modifications for other weather quantities have been developed \citep[see, e.g.,][for an overview]{Hemri&2014, gneiting2014probabilistic, Schefzik&2013}.

Further, postprocessing models allowing to incorporate inter-variable, spatial, or temporal dependence structures have gained increased interest \citep[see, e.g.,][]{moller2013multivariate, baran2015joint, Schefzik&2013, bouallegue2015generation, Berrocal&2007, Feldmann&2015, Kleiber&a2011, Wilks2015, VracFriederichs2015, Schuhen&2012, Pinson2012, Moeller&2016, Hemri&2015}.

A general overview on various aspects of ensemble postprocessing can be found in \citet{Enspostprocessing2018}.

In line with the need for models incorporating dependencies explicitly,
\citet{moller2016probabilistic} introduced a postprocessing model for Gaussian distributed weather quantities (such as temperature) that accounts for dependencies of the individual ensemble forecasts across time. In this regard, the model utilizes the autoregressive information present in the forecast error of the individual raw ensemble forecasts to set up corrected ensemble forecasts as well as a predictive distribution.

The work presented here extends the AR-EMOS model of \citet{moller2016probabilistic}
to be of heteroscedastic (or ``non-homogeneous'') nature, meaning that the
variance parameter of the model varies with the (empirical) ensemble spread.
A postprocessing model using a heteroscedastic variance parameter accounts
for the well-known spread-error correlation \citep{Barker1991,
WhitakerLoughe1998} of forecast ensembles, stating that there is a positive
association between the forecast error (or predictive skill) and the spread
of the ensemble.
Thus, the extended AR-EMOS model incorporates heteroscedasticity in ``two directions'', namely across time (longitudinal) for each individual member, and across the ensemble members (cross-sectional), to account for above mentioned spread-error-correlation. Therefore, the approach allows for features not possible with standard
postprocessing models, such as fitting a predictive distribution based only
on a single ensemble member. This feature is investigated on the basis of an
additional high-resolution forecast added to the ensemble, which is known to improve predictive
performance to a great extend.
While in the original paper of \citet{moller2016probabilistic} the AR-EMOS model was only applied to 24-h ahead ensemble forecasts, in general it is also applicable to (arbitrary) other forecast horizons. In this follow-up work we explain how the AR-EMOS model can be used for other than 24-h ahead forecast horizons and present results on predictive performance.

The development of postprocessing models accounting for specific problems is a quite active area of research, however, not all of the software carrying out model fitting for these recently proposed methods is publicly available.
Prominent examples of postprocessing software implemented under the statistical software environment {\sf R} \citep{R2018} are the state-of-the-art EMOS and BMA models \citep{package-ensembleMOS, package-ensembleBMA}, a recently developed heteroscedastic logistic model for ensemble postprocessing \citep{Messner&2013, Messner&2014, Messner&2016}, implemented in the package \texttt{crch} \citep{package-crch}, and an implementation of verification metrics to assess probabilistic forecasts in the package \texttt{scoringRules} \citep{package-scoringRules}.

In line with the need for publicly available postprocessing software, this follow-up work to the methodology presented in \citet{moller2016probabilistic} is accompanied by an implementation within an \texttt{R} package called \texttt{ensAR} \citep{package-ensAR}, which can currently be installed from the Git repository hosting service GitHub, e.g. by by using the package {\tt devtools} \citep{package-devtools}.

A case study for temperature forecasts of the European Center for Medium Range Weather Forecasts (ECMWF; \citealp{Buizza&2007}) over Germany is carried out to illustrate the performance and properties of the proposed heteroscedastic autoregressive postprocessing model.

\section{Methods}

\subsection{Individual Ensemble Member Postprocessing} \label{sec:armethod}

Suppose that (ensemble) forecasts are initialized at a fixed time point and predict a weather quantity a fixed time step ahead (forecast horizon), which is (for now) not greater than 24 hours. If $t$ denotes the time point (day and hour) for which the forecast is valid, the data consists of forecasts $x_{1}(t), \ldots, x_M(t)$, and a matching weather quantity observation $y(t)$ for $t=1, \ldots,T$. E.g. if a forecast is initialized at 12 UTC and predicts $18$ hours ahead, the forecast is valid at 06 UTC of the following day. Given the initialization time of the forecasts is fixed (which would usually be the case) the data (observation and ensemble members) is a collection of evenly (24 hour) spaced time series referring to the respective validation time point.
Let
\begin{equation} \label{tserror}
Z_{m}(t) := Y(t) - x_{m}(t)
\end{equation}
be the time series of forecast errors of the individual ensemble members $x_m(t)$.
In case $x_m(t)$ would be obtained from an appropriate model for $Y(t)$, the error series $Z_m(t)$ can be assumed to be white noise. However, \citet{moller2016probabilistic} found that the observed individual error series $z_m(t)$ exhibit substantial autoregressive behaviour. The authors propose to utilize this residual autoregressive information to obtain a corrected (AR-adjusted) forecast ensemble and to define a predictive distribution based on this AR-adjusted ensemble forecasts.

In this regard, it is assumed that each $\{Z_{m}(t)\}$ follows an autoregressive process of order $p_m$, denoted by $\text{AR}(p_{m})$, i.e.
$$
Z_{m}(t) - \alpha_{m} = \sum_{j=1}^{p_{m}} \beta_{m,j}[Z_{m}(t-j)- \alpha_{m}] + \varepsilon_{m}(t)\; ,
$$
where $\{\varepsilon_{m}(t)\}$ is white noise with expectation $\text{E}(\varepsilon_{m}(t)) = 0$ and variance \\ $\text{Var}(\varepsilon_{m}(t)) = \sigma_{m}^2$. Then the random variable $Y(t)$ representing the weather quantity can be written as
$$
Y(t) =\widetilde{x}_{m}(t)  + \varepsilon_{m}(t)\; ,
$$
where
$$
\widetilde{x}_{m}(t)  = x_{m}(t) + \alpha_{m} + \sum_{j=1}^{p_{m}} \beta_{m,j}[y(t-j) - x_{m}(t-j) - \alpha_{m}]
$$
can be viewed as a ``corrected'' forecast member for $y(t)$ based on the original ensemble member
$x_{m}(t)$, $x_{m}(t-1)$, \ldots , $x_{m}(t-p_{m})$ at past time points up to and including $t$, and the observation $y(t-1)$, \ldots, $y(t-p_{m})$.

Performing the described procedure for each ensemble member $x_m(t)$ individually yields an ``AR-adjusted'' or ``corrected'' forecast ensemble $\widetilde{x}_1,\ldots,\widetilde{x}_M$.

This approach of obtaining a corrected forecast ensemble rather than a predictive probability distribution/density has a connection to so-called ``member-by-member-postprocssing'' (MBMP), \citealp[see,
e.g.][]{VanSchaeybroeckVannitsem2015, Schefzik2017}). MBMP approaches have gained increased interest, as they retain the dependence structure inherent in the original raw forecast ensemble, while this implicit dependence information is often lost when performing (univariate) postprocessing.

\subsection{Forecast Error Variance} \label{sec:tsproperties}

The variance of the autoregressive process $\{Z_{m}(t)\}$ in Equation \eqref{tserror}, \\
$\text{Var}(Z_{m}(t)) =: \gamma_{m}^2(t)$, is given as
\begin{align}\label{varAR}
\gamma_{m}^2(t) &=\frac{\sigma_{m}^2}{1 - \beta_{m,1} \rho_{m}(1) - \cdots - \beta_{m, p} \rho_{m}(p_{m})},
\end{align}
where $\rho_{m}(k)$ is the autocorrelation function of the process $\{Z_{m}(t)\}$ at lag $k$, see e.g. \citet[Eq. (4.3.31)]{cryer2008time}. In {\sf R}, the autocorrelation function of an ARMA process can be computed with the function {\tt ARMAacf}.

\subsection{Different Forecast Horizons} \label{sec:forchoriz}

If the (ensemble) forecasts are less or equal to 24 hours ahead, it is obvious how to obtain the corrected ensemble forecasts $\widetilde{x}_m$ based on the AR-fit to the error series $Z_m(t)$, as the values $x_{m}(t)$, $x_{m}(t-1)$, \ldots , $x_{m}(t-p_{m})$, as well as $y(t-1)$, \ldots, $y(t-p_{m})$ are readily available at time point $t$. So the required parameters of the AR process can be directly estimated from the (observed) training series $z(t-s), \ldots, z(t-2), z(t-1)$ of length $s$.

However, if the considered forecast horizon is in the interval $(24\text{h}, 48\text{h}]$ and $t$ is any time point (day and hour) for which the forecast(s) are valid, then the observed error $z(t-1)$ is not available, as $y(t-1)$ has not yet been observed. Therefore, a preprocessing step is introduced before the AR-adjusted ensemble can be obtained.
To compensate for the unavailable observed error $z(t-1)$, an AR process is fitted in advance to $z(t-s), \ldots, z(t-3), z(t-2)$, and $z(t-1)$ is predicted from the respective AR model fit.

After predicting $z(t-1)$, the complete error series $z(t-s), \ldots, z(t-2), z(t-1)$ is available again, and the AR-adjusted ensemble can be obtained as described in the previous section. For forecast horizons greater than 48-h, the observed errors at more time steps than $t-1$ are missing (e.g. for 72-h $z(t-1)$ and $z(t-2)$ are missing, and so on). The same procedure can then be applied to predict the missing errors from a fitted AR-model based on the past errors still available.

\subsection{Heteroscedastic Autoregressive Predictive Distribution} \label{sec:heteroscedAR}

\citet{moller2016probabilistic} assume the predictive distribution for $Y(t)$ to be Gaussian, that is
\begin{equation} \label{ARpreddist}
Y(t)|x_1(t),\ldots,x_M(t) \sim \mathcal{N}(\mu(t), \sigma^2(t)),
\end{equation}
where $\mu(t)$ may be a function of the ensemble members and $\sigma^2(t)$ may be a function of the ensemble variance.

To account for the well-known spread-error correlation in ensemble forecasts,
this follow-up work proposes an improved model for the predictive variance
$\sigma^2(t)$ in a similar fashion as the variance term is defined in the EMOS model \citep{gneiting2005calibrated}.

The predictive mean $\mu(t)$ is defined as the average over the AR-adjusted ensemble members
\begin{equation}\label{MeanEmosAR}
\mu(t)= \frac{1}{M} \sum_{m=1}^{M} \widetilde{x}_{m}(t)\; .
\end{equation}
However, an extended variance model is suggested,
combining the estimated dispersion of the error process (longitudinal variation) retrieved solely from
the past of each ensemble member $x_{m}$
with the spread of the
corrected members $\widetilde{x}_{m}$ (cross-sectional variation). To be more specific, the model is given by

\begin{equation}\label{heteroscedVar}
\sigma(t) = w\, \sqrt{\frac{1}{M} \sum_{m=1}^{M}
\gamma_m^2(t)} + (1 - w)\, \sqrt{\widetilde{S}^2(t)}\; ,
\end{equation}
where $\gamma_m^2(t)$ is the variance of the AR process as given in Equation \eqref{varAR},
$\widetilde{S}^2(t)$ is the empirical variance of the AR-corrected ensemble, and $w \in [0,1]$ is
a weight obtained by minimizing the CRPS over an (additional) training period.
A simple (longitudinal) variance model
can be obtained as a special case by setting $w=1$.

The classical EMOS model is also known by the term non-homogeneous regression, due to a variance model that is non-constant (non-homogeneous/hetero\-scedastic) with respect to the spread in the ensemble. Therefore, the modification of the AR-EMOS model with the variance model \eqref{heteroscedVar} incorporating the ensemble spread, is called heteroscedastic AR-EMOS - following the EMOS nomenclature.
As the AR-EMOS method is based on a time-series model, and does not only consider the variation in the ensemble spread (cross-sectional part), but also the variation across time (longitudinal part), the more general (and in statistics more common) term heteroscedastic is used rather than the term non-homogeneous.

\subsection{Postprocessing a Single Forecast}

When there exists a single distinguished forecast $x_{\ast}(t)$, e.g. the high-resolution forecast $x_{\text{hres}}(t)$ described below, it is still possible to obtain a corresponding predictive distribution by the described AR-EMOS method.

The parameters $\mu_{\ast}(t)$ and $\sigma_{\ast}(t)$  are in principle estimated in the same way as those corresponding to the regular members. However, the mean \eqref{MeanEmosAR} and the longitudinal part of the variance formula  \eqref{heteroscedVar} reduce to a single summand, and the second part of the variance model in Equation \eqref{heteroscedVar} becomes zero. Nonetheless, the variance $\gamma_{\ast}^2$ of the  error series corresponding to the individual member $x_{\ast}(t)$ can still be computed - on the basis of past values and the AR-fit.
Thus, the original AR approach and its refined version presented here both allow to estimate the variance parameter and fit a predictive distribution based only on a single forecast. This is a new and useful feature not available in other state-of-the-art postprocessing models.

\subsection{High-Resolution Forecast} \label{sec:hresForc}

In this section we describe how the high-resolution forecast known to improve predictive performance \citep[see, e.g.][]{Kann&2009, gneiting2014calibration, ECMWFGuide2015} can be included into the AR-EMOS postprocessing model. We call the AR-EMOS model including the additional forecast an extended model.

Let again $x_{1}(t),\ldots, x_{M}(t)$ denote the
forecast ensemble. However, this time the $M$ members comprise of the regular (exchangeable) forecasts described in Section \ref{sec:armethod} and the additional high-resolution forecast $x_{\text{hres}}(t)$. That means the (total) number $M$ of forecasts utilized is actually increased by one.
As described in Section \ref{sec:armethod} it is again assumed that the forecast errors $Z_m(t)$, $m=1,\ldots,M$ follow an $\text{AR}(p_m)$ process, yielding the AR-corrected ensemble $\widetilde{x}_{1}(t),\ldots, \widetilde{x}_{M}(t)$, $\widetilde{x}_{\text{hres}}(t)$, which is the basis for estimating mean and variance of the predictive distribution.

As the high-resolution forecast has somewhat different properties than the regular ensemble members, an apparent approach may be to treat them as two different (exchangeable) groups with respect to the parameters of the predictive distribution. This course of action is quite common in ensemble post-processing models, ensemble members belonging to a certain (exchangeable) group are assumed to share the same coefficients in the model \citep{gneiting2014calibration}.

To account for the above mentioned groups, each parameter in the AR-EMOS model is defined as (equally weighted) sum of the respective group-wise parameters,  i.e.,
\begin{align*}
\mu(t)&=\frac{1}{2} \,(\mu_{\text{ens}}(t) + \mu_{\text{hres}}(t))\\
\sigma(t)&=\frac{1}{2} \,(\sigma_{\text{ens}}(t) + \sigma_{\text{hres}}(t))
\end{align*}

Here, $\mu_{\text{ens}}(t)$ is estimated as already stated in Equation \eqref{MeanEmosAR}.
The parameters $\mu_{\text{hres}}(t)$ and $\sigma_{\text{hres}}(t)$ corresponding to the high-resolution forecast are  estimated as described in the precious subsection.

Assigning each of the group-specific parameters fixed and equal weights is an initial and relatively straightforward approach for demonstrating the general idea. In the case study it will be shown that this simple version already yields good results for predictive performance.

Of course this rather simple method can be modified to be more data driven, that is using weights for the group-specific parameters directly estimated from data, e.g. by minimum CRPS estimation.
Furthermore, the approach for two groups (regular ensemble members and high-resolution forecast) can be generalized to include multiple (exchangeable) groups, which also need not necessarily contribute equally to predictive performance, which makes it reasonable to estimate the weights for the group-specific parameters from data. 

One possibility for a more general definition of an AR-EMOS group model can be accomplished for example by combining (group-wise) predictive distributions with the spread-adjusted linear pool already employed by  \citet{moller2016probabilistic} and shortly described in the following subsection.

\subsection{Combination of Predictive Distributions}\label{sec:combdist}

\citet{moller2016probabilistic} proposed to combine the predictive distribution of classical EMOS and AR-EMOS in a spread-adjusted linear pool (SLP, \citealp{gneiting2013combining}).
For the special case of combining $n=2$ predictive distributions, the SLP combination has cumulative distribution function (CDF)
$$
F(x) = w_{1} G_{1}(x) + w_{2} G_{2}(x), \quad G_{l}(x) =
\Phi\left(\frac{x- \mu_{l}}{\sigma_{l} c}\right)\; ,
$$
$l=1,2$, where $w_{1}$ is a non-negative weight parameter, $w_{2}= 1- w_{1}$, and $c$ is a strictly positive
spread adjustment parameter. Here, $\phi$ and $\Phi$ denote the probability density function (PDF) and the
cumulative distribution function (CDF)
of the standard normal distribution, respectively. The two distributions $G_1$ and $G_2$ can be fitted separately by postprocessing models of choice, and the weights are obtained by minimizing a verification score (specifically the CRPS) over a training period, for fixed and given $G_1, G_2$.

The original approach was proposed with the aim to improve predictive performance by combining two predictive distributions coming from different sources \citep[see also][]{gneiting2013combining}.
This approach can directly be extended to a (finite) number $n>2$ of predictive distributions.

\section{Tools to Asses Predictive Performance} \label{VeriMetrics}

\subsection{Scoring Rules}

A common tool to assess the quality of probabilistic forecasts are scoring rules. They assign a scalar to a pair ($y$,$F$), where $y$ is the verifying observation and $F$ the forecasting distribution \citep{Gneiting2011,GneitingRaftery2007, GneitingBalabdaouiRaftery2007}.
Scoring rules are negatively orientated such that smaller values indicate better performance.

A well-known and popular score is the continuous ranked probability score (CRPS), assessing calibration and sharpness simultaneously. For a predictive distribution $F(y)$ and observation
$y_{\text{obs}}$ the CRPS is given as
\begin{equation}\label{crps}
\text{CRPS}(F, y_{\text{obs}}) = \int_{-\infty}^{\infty}
\left\{F(y) - \boldsymbol{1}(y\geq y_{\text{obs}})\right\}^{2}
\, \text{d} y\; ,
\end{equation}
where $\boldsymbol{1}(y\geq y_{\text{obs}})$ equals $1$ if $y\geq y_{\text{obs}}$ and
$0$ otherwise, see also \citet[Sect. 8.5.1]{wilks2011statistical}.

If $\mu_{F}$ and $\sigma_{F}^2$ denote the mean and variance of $F$, the
\citet{dawid1999coherent} score is given by
\begin{equation}
\text{DSS}(F, y_{\text{obs}}) = \frac{(y_{\text{obs}} - \mu_{F})^2}{\sigma_{F}^2} + 2\log \sigma_{F}\; ,
\end{equation}
see also \citet{gneiting2014probabilistic}.

\subsection{Visual Assessment}

To visually assess calibration of a probabilistic forecast,
verification rank histograms and PIT histograms are employed
\citep{wilks2011statistical}.

Here, the verification rank histogram (VRH) or Talagrand diagram is used to
assess a forecast ensemble $x_1,\ldots,x_M$. It can be obtained by computing
the rank of the observation $y$ within the ensemble (for each forecast case).
If the ensemble members $x_1,\ldots, x_M$ and the observation $y$ are
statistically indistinguishable (exchangeable), the rank of the observation
with respect to the ensemble members has a discrete uniform distribution on
$\{1,\ldots,M+1\}$. The VRH then plots the empirical frequency of the
observation ranks.

To assess calibration of a full predictive probability distribution, the
frequencies of the Probability Integral Transform (PIT) values are plotted in
equidistant bins. An observation $y$ can be interpreted as a random sample
from the ``true'' distribution $F$ for the respective weather quantity. If
the predicted distribution $F_0$ is identical to $F$, then $p=F_0(y)$ can be
considered as realization of a uniformly distributed random variable on
$[0,1]$ and the plot of the frequencies of the PIT values $p$ results in a
uniform histogram.

\subsection{Further Verification Measures}

The variance of the PIT values provides further information on the dispersion
properties of the predictive distribution, a neutral dispersion being
indicated by a variance equal to $\frac{1}{12}=0.0833$, the variance of the
uniform distribution on $[0,1]$, see \citep{gneiting2013combining}.

The root mean variance (RMV) is used as a sharpness measure of predictive
probability distributions. A main principle of probabilistic forecasting is
``maximizing the sharpness of the predictive distribution subject to
calibration'' \citet[see e.g.][]{gneiting2014probabilistic}, therefore the
sharpness should be investigated in conjunction with the calibration.

\subsection{Testing for Improvement in Predictive Performance}

The statistical relevance of improvement in the verification scores may be investigated by testing for equal predictive
performance of the two considered methods with the Diebold-Mariano test for time series, see \citet{gneiting2014probabilistic}.

Let $s_{1}(t), s_{2}(t) $ denote the time series of score values obtained from two
competing methods as for example EMOS and AR-EMOS, for a verification period
of length $T$ (say). Then the large-sample standard normal test statistic
adapted from \citet{diebold1995comparing} is given as
$$ S = \sqrt{T}
\frac{\overline{d}}{\sqrt{\displaystyle\sum_{\tau = - (h-1)}^{h-1}
\widehat{\gamma}_{d}(\tau)}}\; , $$
where
$$ \overline{d} = \frac{1}{T}
\sum_{t=1}^{T} d(t), \quad d(t) = s_{1}(t) - s_{2}(t) \; , $$
is the average
CRPS differential and $$ \widehat{\gamma}_{d}(\tau) = \frac{1}{T} \sum_{t =
|\tau|+1}^{T} \, (d(t) - \overline{d})(d(t-|\tau|) - \overline{d}) $$ are the
empirical autocovariances.

\section{Application to ECMWF temperature forecasts}

\subsection{Data Description and Data Preprocessing} \label{sec:data}

The considered data for our case study contains the forecast ensemble with 50 members of the European Centre for Medium-Range Weather Forecasts (ECMWF, see e.g. \citealp{molteni1996ecmwf, Buizza&2007}). The data consists of $24$-h, $48$-h and $72$-h ahead forecasts initialized at 12 UTC for 2-m surface temperature in Germany along with the verifying observations at 187 different stations in the time period ranging from 2002-01-01 to 2014-03-20, see also \citep{Hemri&2014}. In addition, there is one high-resolution forecast and one control forecast.

For the application of time series methods it is of importance to investigate whether the dates appear in chronological order and if some dates are missing.
From the full data set with 187 stations only those stations are retained, which do not reveal NA gaps longer than 1. There are 76 stations which do match this rather strict specification. The new data set still contains missing values.

\begin{figure}
\centering
\includegraphics[width=70mm]{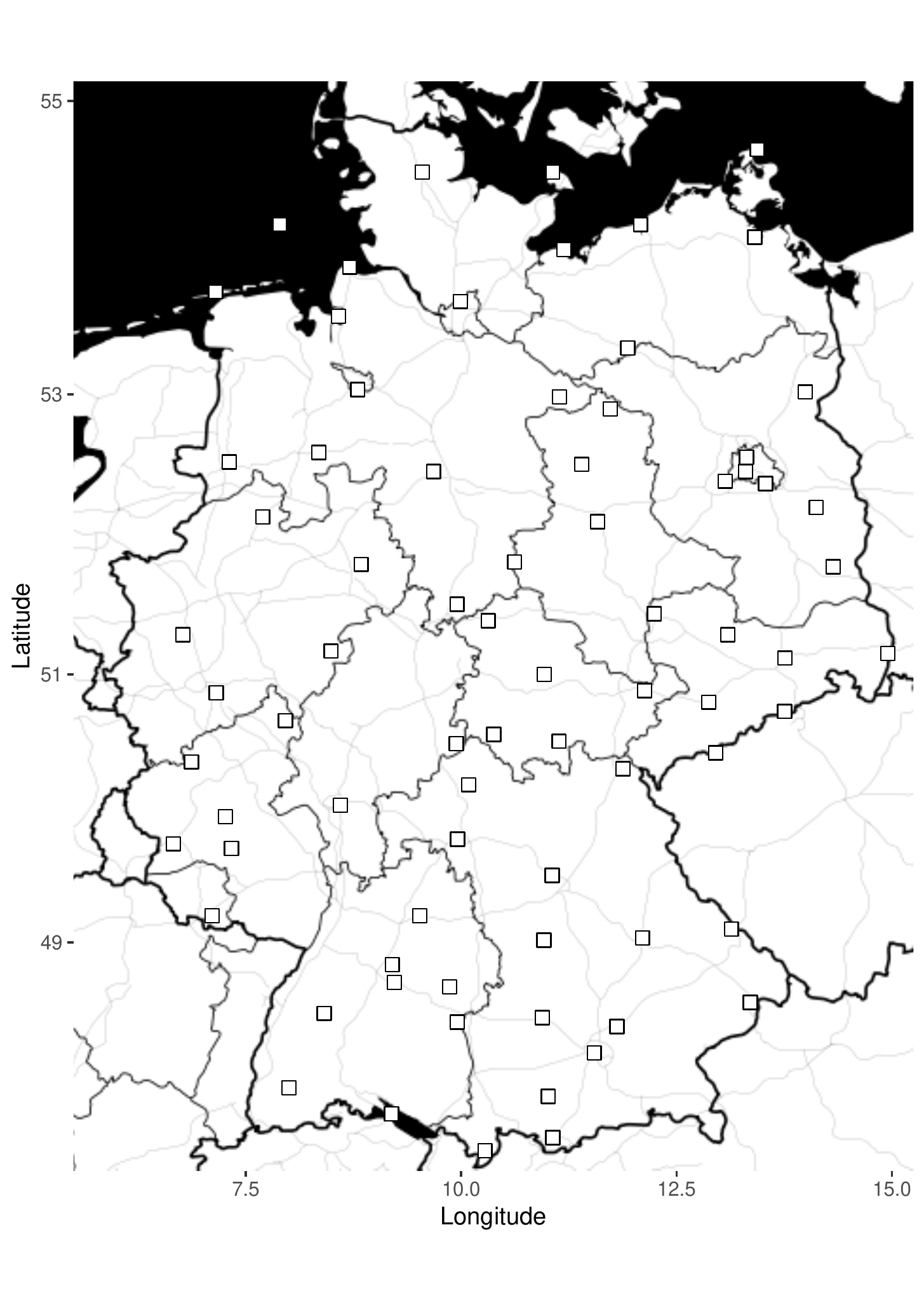}
\caption[76 stations in Germany]{76 stations in Germany, chosen for admitting a modest occurrence of missing values in the raw data set.} \label{StationsGermany}
\end{figure}

The remaining missing values can in {\sf R} be replaced by values obtained from linear interpolation using the function {\tt na.approx} from the package {\tt zoo} \citep{package-zoo}.

Figure \ref{StationsGermany} shows the 76 stations retained for the subsequent analysis, where the station map was produced with the {\sf R} package {\tt ggmap} \citep{package_ggmap}. In the subsequent analysis the station Magdeburg in Eastern Germany will be considered for illustration purposes in the case study.

To find possible outliers in the data, the
test statistic from \citet{chang1988estimation} for detecting additive outliers is applied, being implemented as function {\tt detectAO} in the {\sf R} package {\tt
TSA} \citep{package-TSA}. It requires the fit of an ARIMA model to the series, which can e.g. be achieved by the function {\tt auto.arima} from the {\sf R} package {\tt forecast} \citep{package-forecast}.

The analysis reveals one indicated suspicious observation.
From further inspection, the detected value of $30.6\,^{\circ}\mathrm{C}$ at 2011-08-26 appears a bit too high with respect to preceding and succeeding observations, but may nonetheless be a true value.
Thus, it is concluded that the observation should not be altered or removed.

In order to find strong outlying observations, the significance level in {\tt detectAO} is put to the very small value $\alpha = 0.00001$. By applying the above procedure to each station, it is found that 6 stations reveal suspicious values.
From these, only two observations are removed from the data set. At date
2003-01-20 and station Hannover the temperature observation is $-90.8\,^{\circ}\mathrm{C}$, which is clearly impossible. At date
2002-11-23 and station Nuernberg, the temperature observation is $-20.1\,^{\circ}\mathrm{C}$, which is very unusual with respect to preceding and succeeding temperature values and, in addition, is by far the smallest value in the complete series.

Removal is done by setting the outlying value to NA and then applying linear interpolation.

\subsection{Comparison of EMOS and Heteroscedastic AR-EMOS}\label{sec:modcomparison}

First, the state-of-the-art EMOS model is compared to the heteroscedastic AR-EMOS model (called AR-EMOS in the following) presented in Section \ref{sec:heteroscedAR}, where each model is based on the 50 regular ECMWF 24-h ahead ensemble forecasts. The parameters of the postprocessing models are estimated station-wise, based only on the data available for the respective station (so-called local approach).

A first analysis investigates the performance of the different models at the station Magdeburg in Eastern Germany. Then, in a second step the analysis is carried out for all 76 stations in the data set.

\subsubsection{Settings}

To fit the AR-EMOS model presented in Section \ref{sec:heteroscedAR} (with the basic methodology in Section \ref{sec:armethod}), first the model parameters $\alpha_{m}, \beta_{m,1}, \ldots, \beta_{m,p_{m}}$ for each member $x_m$ are estimated by fitting an $\text{AR}(p_{m})$ process to the observed error series $\{z_{m}\}$ from a
training period by Yule-Walker estimation as carried out by
the function {\tt ar}, see also \citet[Section 3.6]{shumway2006time}.
The order $p_{m}$ is automatically selected by a modified Akaike information criterion (AIC) proceeding as if the required variance estimate were obtained by maximum likelihood, cf. \citet[function {\tt ar}]{R2018}.

This procedure is invoked with a default training length of $90$ days, which has been found appropriate by \citet{moller2016probabilistic}.

To obtain the predictive mean $\mu(t)$ and standard deviation $\sigma(t)$ (as defined in Equation \eqref{heteroscedVar})
a second training period is required to estimate the weight parameter $w$ used in the heteroscedastic variance model \eqref{heteroscedVar} such that the average CRPS with respect to the predictive distribution $N(\mu(t), \sigma(t)^2)$ is minimized for the training period. In the case study following later, the default for the additional training period is set to 30 days length.

The EMOS model is fitted with the {\sf R} package {\tt ensembleMOS}. Estimating parameters of EMOS usually requires a training period of length between $20$ to $40$ days \citep[Sect. 4]{gneiting2014calibration}, where in the subsequent study $30$ days are chosen.

\begin{figure}
\centering
\includegraphics[width=130mm, height=87mm]{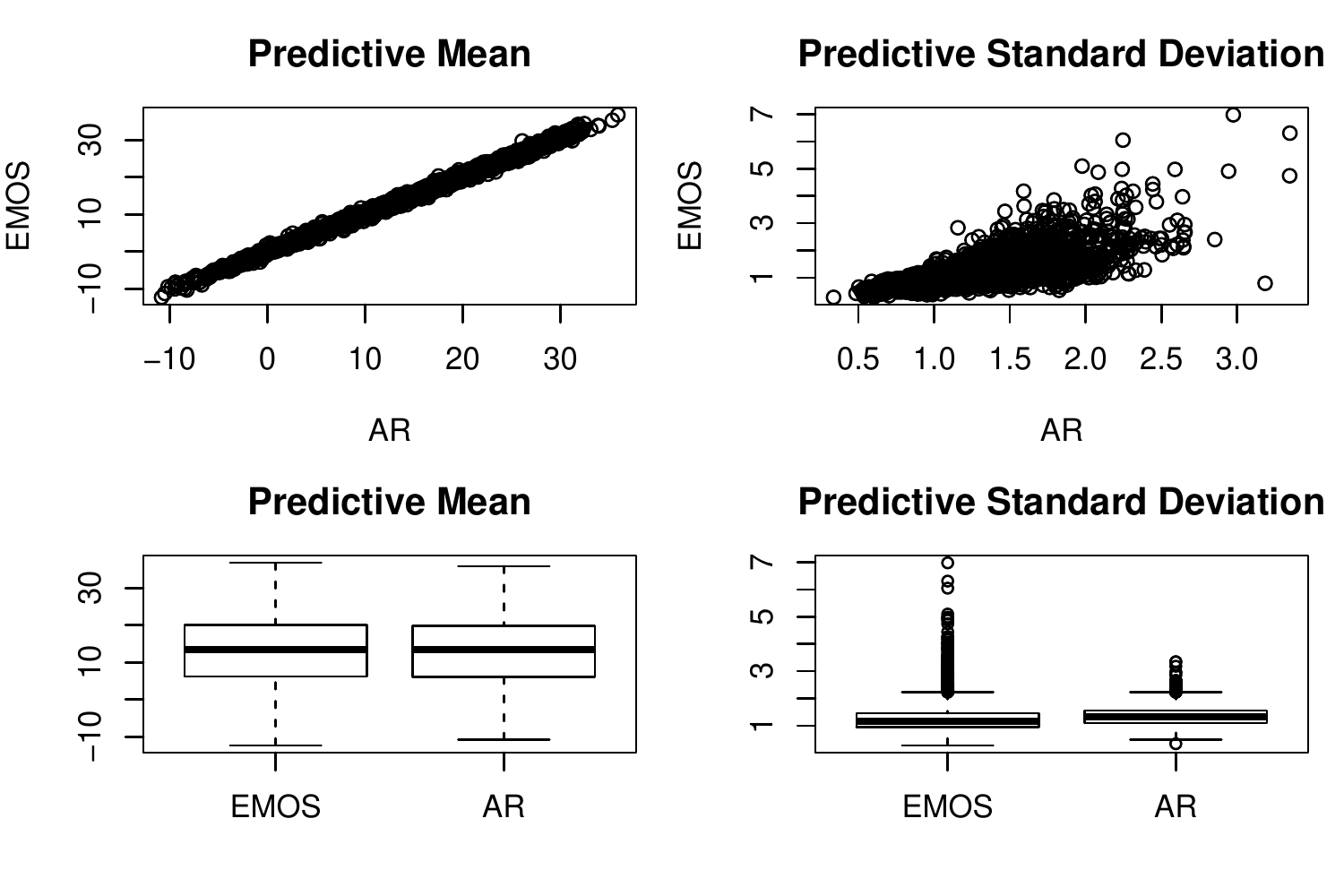}
\caption[Comparison of predictive mean and predictive standard deviation obtained by EMOS and AR-EMOS]{Comparison of predictive mean and predictive standard deviation obtained by EMOS and (heteroscedastic) AR-EMOS}\label{fig:MeanSdMD}
\end{figure}

\subsubsection{Comparison at a Single Station}

The plots displayed in Figure \ref{fig:MeanSdMD} show a comparison of the predictive mean and standard deviation of the EMOS and (heteroscedastic) AR-EMOS predictive distribution for Magdeburg. The predictive means of both methods exhibit a strong relationship (squared correlation coefficient equal to 0.9963), while the predictive standard deviations differ to a certain extent (squared correlation coefficient equal to 0.5142). So, although, AR-EMOS proceeds in a quite different way to estimate the predictive mean, the result does apparently not differ from EMOS very much. The different approaches to estimate the variance obviously also yield different results, the AR-EMOS standard deviations have a tendency to be smaller than those of EMOS. The Boxplot for the stand deviations also show that for EMOS there is much more variation in the estimated standard deviations than for AR-EMOS.

Table \ref{tab:verifMD} presents the CRPS, the DSS, the root mean variance (RMV) and the PIT variance for both methods at station Magdeburg. The top row in Figure \ref{fig:PITAll} additionally shows the PIT histograms of both methods at station Magdeburg.

\renewcommand{\arraystretch}{1.2}
\setlength{\tabcolsep}{1.2mm}
\begin{table}[!ht]\centering
\begin{tabular}{lcccc}
\toprule[0.09 em]
~ & CRPS & DSS & RMV & $\textup{Var}$(PIT) \\
\midrule
EMOS & 0.8415 & 2.0918 & 1.3670 & 0.0946 \\
AR-EMOS & 0.8309 & 1.9149 & 1.3825 & 0.0876 \\
\bottomrule[0.09 em]
\end{tabular}
\caption{Verification metrics of EMOS and (heteroscedastic) AR-EMOS for station Magdeburg aggregated over 4341 verification dates \label{tab:verifMD}}
\end{table}

The PIT values of both models have a variance greater than $1/12=0.0833$, indicating underdispersion of the predictive distributions. This underdispersion is visible in the respective PIT histograms as well. However, the PIT variance of AR-EMOS is much closer to $1/12$ than the PIT variance of EMOS. When looking at the PIT histograms, the EMOS histogram indicates a slightly more pronounced bin for small PIT values, indicating a stronger forecast bias.
On the contrary, the EMOS predictive distribution is slightly sharper than the AR-EMOS one, however obviously at an expense in dispersion accuracy.

When looking at the verification scores providing an overall judgment on predictive performance (CRPS, DSS), we can further conclude that AR-EMOS performs better than EMOS with respect to the average CRPS as well as to the average DSS at station Magdeburg.

As the difference in CRPS values is relatively small, we investigate whether AR-EMOS provides a statistically significant improvement in CRPS over state-of-the-art EMOS, by a one-sided Diebold-Mariano test for the alternative \\
$$H_1: CRPS_{\text{AR-EMOS}} < CRPS_{\text{EMOS}}\; .$$
For station Magdeburg, the resulting p-value is given as 0.01722. Thus, the test shows that the CRPS values of AR-EMOS are (on average) indeed significantly smaller than those of EMOS.

\subsubsection{Comparison at all Stations}

In a second step the analysis performed for a single example station is carried out for all 76 stations in the data set, and the results are aggregated. For this, a local approach is used, that is, the considered models are estimated at each station individually, resulting in location-specific model parameters based only on the data of a specific station.

\begin{figure}
\centering
\includegraphics[width=115mm, height=92mm]{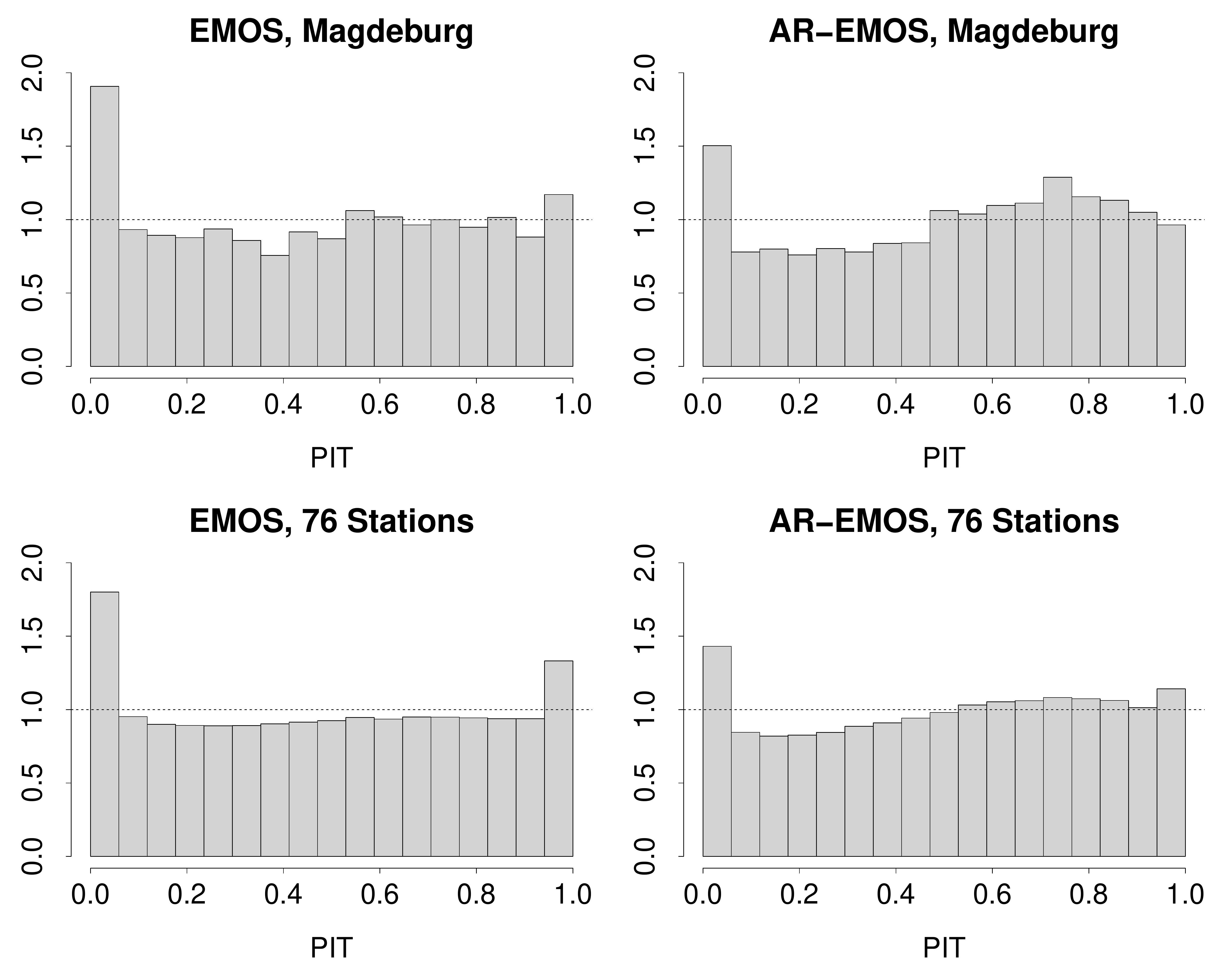}
\caption{PIT histograms of EMOS and (heteroscedastic) AR-EMOS aggregated over 4341 verification days (top row), as well as PIT histograms of both methods aggregated over $4341 \times 76$ verification cases (dates and stations, bottom row) \label{fig:PITAll}}
\end{figure}

When aggregating over all 76 stations, for each station 4341 verification days are available, yielding in total 329916 forecast cases to aggregate over. Table \ref{tab:verifAll} shows the resulting verification metrics for EMOS and AR-EMOS.

\renewcommand{\arraystretch}{1.2}
\setlength{\tabcolsep}{1.2mm}
\begin{table}[!ht]\centering
\begin{tabular}{lcccc}
\toprule[0.09 em]
~ & CRPS & DSS & RMV & $\textup{Var}$(PIT)\\
\midrule
EMOS & 0.9057 & 2.1523 & 1.4907 & 0.0954 \\
AR-EMOS & 0.9033 & 2.0129 & 1.5322 & 0.0883 \\
\bottomrule[0.09 em]
\end{tabular}
\caption{Verification metrics of EMOS and (heteroscedastic) AR-EMOS aggregated over $4341 \times 76$ verification cases (dates and stations) \label{tab:verifAll}}
\end{table}

The verification scores indicate that the predictive distribution of AR-EMOS has comparable but slightly different properties than EMOS, similar to the comparison at a single station. While CRPS and DSS of AR-EMOS are (slightly) smaller than those of EMOS, the RMV is slightly larger, indicating the AR-EMOS predictive distribution to be less sharp than its EMOS counterpart. However, the sharper distribution of EMOS comes at the expense of calibration, the PIT variance of EMOS is much larger than 1/12, indicating underdispersion, while the PIT variance of AR-EMOS is relatively close to $1/12$, indicating a distribution with dispersion properties close to neutral dispersion. These observations are consistent with the PIT histograms in (the bottom row of) Figure \ref{fig:PITAll}, where the EMOS PIT histogram clearly exhibits a U-shape, with indicates underdispersion, while the U-shape is much less pronounced in the AR-EMOS PIT histogram.

To find statistical evidence about the significance of the difference in predictive performance between the two methods, the one-sided Diebold-Mariano test is conducted again, this time for the CRPS time series at each of the 76 stations individually.

\begin{figure}
\centering
\includegraphics[width=145mm, height=102mm]{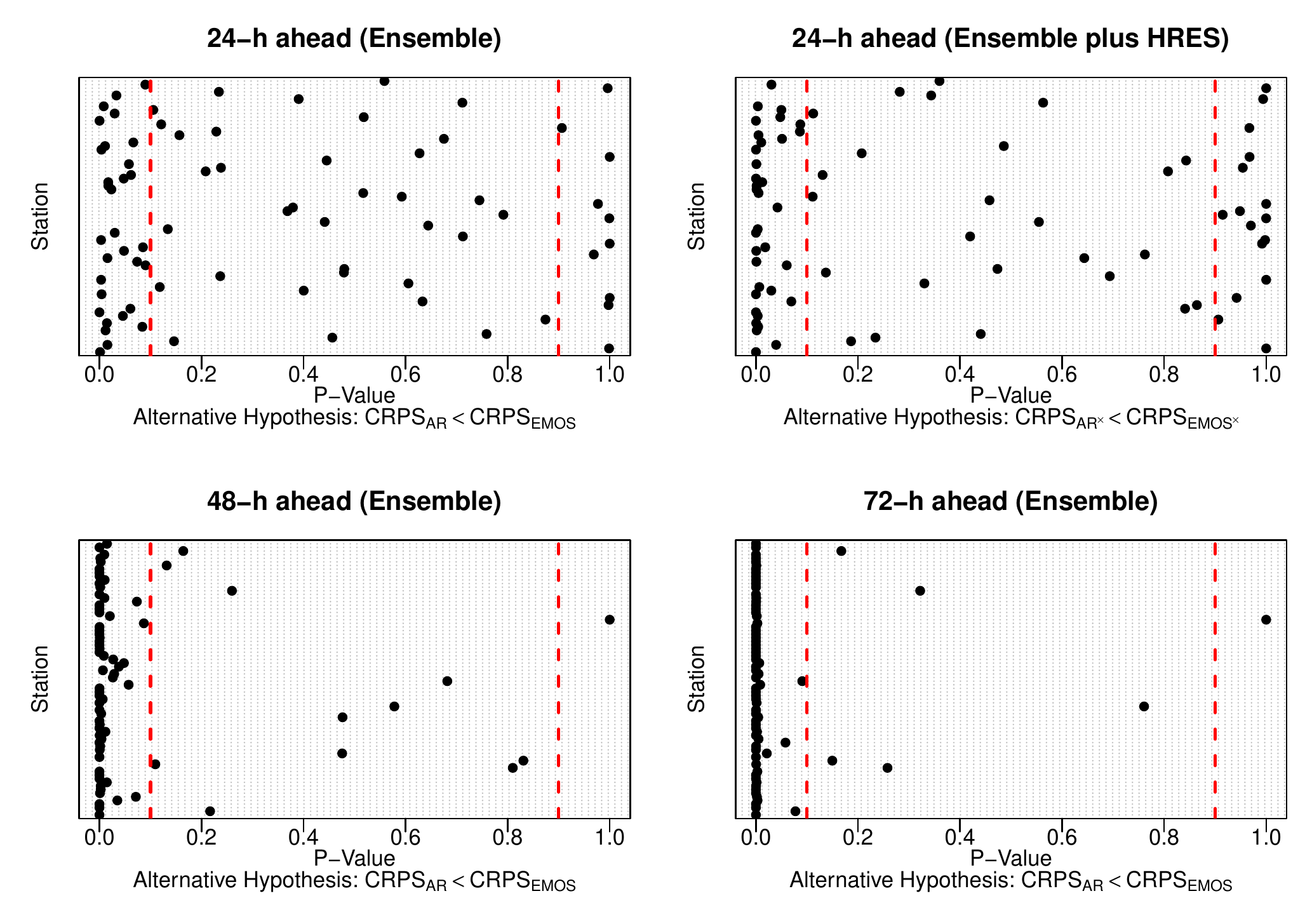}
\caption[P-values of the Diebold-Mariano one-sided test for equal predictive performance for 76 stations]{Station-wise p-values of the Diebold-Mariano one-sided test comparing EMOS and (heteroscedastic) AR-EMOS.}\label{fig:DMtest}
\end{figure}

The upper left panel of Figure \ref{fig:DMtest} shows the station-wise p-values of the Diebold-Mariano test for EMOS vs. AR-EMOS. Small p-values give statistical evidence for the alternative hypothesis that the values of the AR-EMOS CRPS series are on average smaller than the values of the EMOS CRPS series, thus indicating superior performance of AR-EMOS compared to EMOS.
At 31 stations the p-value is $\leq 0.1$, thus indicating superior performance of AR-EMOS, see the left vertical red dashed line in (the upper left panel of) Figure \ref{fig:DMtest}.

Large p-values give statistical evidence for the converse alternative hypothesis, thus indicating inferior performance of AR-EMOS compared to EMOS. However, there are only 10 stations with p-value greater than 0.9, see right vertical dashed red line.

Thus, for a large portion of the considered stations, AR-EMOS exhibits significantly better predictive performance than EMOS in terms of the CRPS, while EMOS is significantly superior to AR-EMOS only at a few stations.

\subsection{Incorporating the High Resolution Forecast with Group Approach}\label{sec:hres}

The ECMWF ensemble also comprises a single high-resolution run, whose importance for statistical postprocessing is described e.g. by \citet{gneiting2014calibration}.
As indicated in Section \ref{sec:hresForc}, extended postprocessing models are now considered, which include the high-resolution forecast.

For the ECMWF data considered here, there are 50 exchangeable (that is statistically indistinguishable) forecast members forming one group, while the high-resolution forecast is regarded as second group due to its different properties.

Within the \texttt{ensembleMOS} package, the group membership of each ensemble forecast can be directly specified. As described in Section \ref{sec:hresForc}, a straightforward ad-hoc way to implement a 2-group AR-EMOS model is to represent the model parameters as a sum of the two group specific parameters and assign the group-specific parameters (fixed) equal weight.

Furthermore, the SLP combination of EMOS and AR-EMOS proposed in \citet{moller2016probabilistic}
is revisited. However, in contrast to the original analysis, here the SLP combination of EMOS
and AR-EMOS based on the 50 exchangeable members \textit{and} the additional high-resolution
forecast is investigated (which we called extended models, denoted by EMOS$^{\ast}$ and AR-EMOS$^{\ast}$, respectively).

As additional training data is needed to estimate the weights in the SLP combination, the final number of verification days considered differs from the above analyses comparing only EMOS and AR-EMOS.
Here, the results at the station Magdeburg are aggregated over 4251 verification days. When aggregating over all 76 stations (each with 4251 verification days) as well, the results are based on 323076 forecast cases in total.

Results for verification scores at the station Magdeburg are presented in Table \ref{tab:verifSLPMD}. It is clearly visible that in terms of CRPS and DSS AR-EMOS$^{\ast}$ improves over EMOS$^{\ast}$ to a large extent. The improvement is much more pronounced than in the case where the high-resolution forecast was not incorporated into both models. The SLP$^{\ast}$ combination of the two models improves both scores even more in comparison to EMOS$^{\ast}$.

Concerning sharpness as measured by the RMV the AR-EMOS$^{\ast}$ model yields the sharpest predictive distribution, with a PIT variance extremely close to $1/12$ at the same time. EMOS$^{\ast}$ and the SLP combination of both models are less sharp (they exhibit nearly the same level of sharpness), however, while EMOS$^{\ast}$ as a PIT variance indication underdispersion (larger $1/12$), the SLP$^{\ast}$ combination also has a PIT variance close to neutral dispersion. Therefore, at the station Magdeburg, the sharpness-calibration properties of AR-EMOS$^{\ast}$ seem to be extremely appropriate and better than those of the other predictive distributions.

\renewcommand{\arraystretch}{1.2}
\setlength{\tabcolsep}{1.2mm}
\begin{table}[!ht]\centering
\begin{tabular}{lcccc}
\toprule[0.09 em]
~ & CRPS & DSS & RMV & $\textup{Var}$(PIT)\\
\midrule
EMOS$^{\ast}$ & 0.8223 & 2.0304 & 1.3714 & 0.0908 \\
AR-EMOS$^{\ast}$ & 0.8097 & 1.8404 & 1.3663 & 0.0849\\
SLP$^{\ast}$ & 0.8000 & 1.9043 & 1.3965 & 0.0854 \\
\bottomrule[0.09 em]
\end{tabular}
\caption{Verification metrics of EMOS$^{\ast}$, (heteroscedastic) AR-EMOS$^{\ast}$, and SLP combination SLP$^{\ast}$of both, aggregated over 4251 verification dates at station Magdeburg \label{tab:verifSLPMD}}
\end{table}

Although the improvement in CRPS of AR-EMOS$^{\ast}$ compared to EMOS$^{\ast}$ is much more obvious as in the respective analysis at Magdeburg presented in Table \ref{tab:verifMD}, a one-sided Diebold-Mariano test for $H_1: CRPS_{\text{AR-EMOS$^{\ast}$}} < CRPS_{\text{EMOS$^{\ast}$}}$ at Magdeburg is performed to investigate the significance of the improvement. The resulting p-value is $0.01233$, showing that AR-EMOS$^{\ast}$ is indeed performing significantly better than EMOS$^{\ast}$ at Magdeburg in terms of CRPS.

\subsubsection{Illustration of Predictive Distributions}

\begin{figure}
\centering
\includegraphics[width=115mm, height=81mm]{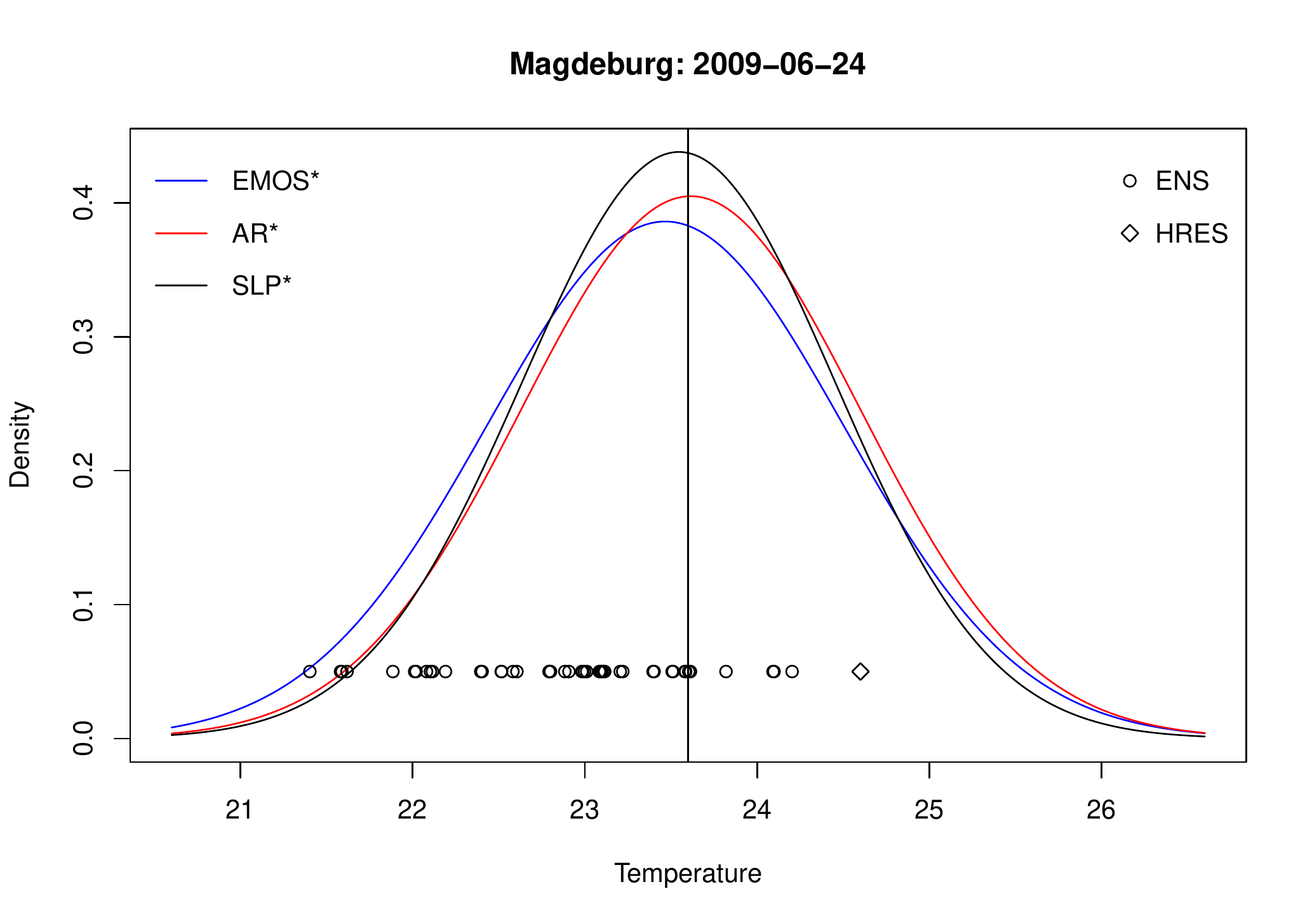}
\caption{Predictive pdfs based on the 50 ensemble members and the high-resolution forecast for Magedburg, valid at date 2009-06-24, along with the respective observation (vertical line)\label{fig:predPDF}}
\end{figure}

For a visual illustration the three predictive probability density functions (PDFs) corresponding to EMOS$^{\ast}$, AR-EMOS$^{\ast}$, and SLP$^{\ast}$ (two normals and a mixture of two normals as described in Section \ref{sec:combdist}) are graphically displayed for the randomly chosen forecast day 2009-06-24, see Figure \ref{fig:predPDF}.

Here, it is clearly visible that the PDF of AR-EMOS$^{\ast}$ and the SLP$^{\ast}$
combination are more centered around the observation in comparison to the PDF of EMOS,indicating a higher consistency of the respective distributions with the true observation, thus improved calibration. This observation is in line with the verification results presented in Table \ref{tab:verifSLPMD}.

\subsubsection{Comparison of all Stations}

Next, the above described comparison of EMOS$^{\ast}$, AR-EMOS$^{\ast}$ and SLP$^{\ast}$ is conducted for all 76 stations, where again the models are estimated station-wise. Due to the need for additional training data for the SLP combination, the number of verification cases considered again differs from the analysis presented in Table \ref{tab:verifAll} as mentioned for the station Magdeburg. Here, the aggregation in Table \ref{tab:verifSLPAll} is performed over 323076 verification cases (4251 verification dates at each of the 76 stations). The aggregated verification metrics show that AR-EMOS$^{\ast}$ performs better than EMOS$^{\ast}$ with respect to the CRPS and DSS. With respect to the CRPS the SLP combination performs best, with respect to DSS it also performs clearly better than EMOS$^{\ast}$.
EMOS$^{\ast}$ has the sharpest predictive distribution in terms of the RMV, while AR-EMOS$^{\ast}$ and SLP${\ast}$ exhibit a similar level of sharpness.
However, the PIT variance of AR-EMOS$^{\ast}$ is much closer to that of neutral dispersion than EMOS$^{\ast}$, having a PIT variance indicating underdispersion.

\renewcommand{\arraystretch}{1.2}
\setlength{\tabcolsep}{1.2mm}
\begin{table}[!ht]\centering
\begin{tabular}{lcccc}
\toprule[0.09 em]
~ & CRPS & DSS & RMV & $\textup{Var}$(PIT)\\
\midrule
EMOS$^{\ast}$ & 0.8712 & 2.1412 & 1.4250 & 0.0931 \\
AR-EMOS$^{\ast}$ & 0.8685 & 1.9270 & 1.4950 & 0.0854 \\
SLP${\ast}$ & 0.8460 & 1.9350 & 1.5031 & 0.0860 \\
\bottomrule[0.09 em]
\end{tabular}
\caption{Verification metrics of EMOS$^{\ast}$, (heteroscedastic) AR-EMOS$^{\ast}$, and SLP combination SLP$^{\ast}$ of both,
aggregated over $4251 \times 76$ verification cases (dates and stations) \label{tab:verifSLPAll}}
\end{table}

Again, station-wise p-values of the one-sided Diebold-Mariano test for the CRPS series are computed to investigate whether the improvement in CRPS is significant. The upper right panel of Figure \ref{fig:DMtest} shows the resulting p-values at all 76 stations for the one-sided Diebold-Mariano test for the alternative $H_1: CRPS_{\text{AR-EMOS$^{\ast}$}} < CRPS_{\text{EMOS$^{\ast}$}}$.

Again, the dashed red lines denote the significance levels 0.1 and 0.9, respectively. When incorporating the high-resolution forecast into the models, the number of stations where AR-EMOS$^{\ast}$ performs significantly better (at significance level 0.1) than EMOS$^{\ast}$ increases to 41. At 16 stations the p-value is greater than 0.9, indicating that EMOS$^{\ast}$ performs better than AR-EMOS$^{\ast}$.

\subsection{Postprocessing for Higher Forecast Horizons}\label{sec:higherForcHoriz}

Finally we investigate the performance of the heteroscedastic AR-EMOS model for higher forecast horizons. In many applications, typically 24-h ahead forecasts are investigated, but often higher forecast horizons are not considered.
To illustrate the effect, we present results for 48-h and 72-h ahead forecasts.

In Section \ref{sec:forchoriz} the procedure for applying AR-EMOS to forecast horizons greater than 24-h is explained. EMOS is capable of dealing with other than 24-h ahead forecasts as well, and the forecast horizon considered can explicitly specified within the \texttt{ensembleMOS} package.

Table \ref{tab:verifAll48h72h} shows the verification metrics for EMOS and AR-EMOS, based on 48-h and 72-h ahead ensemble forecasts.
For 48-h ahead forecasts the verification metrics and PIT histograms are based on a total of 329840 verification cases (4340 verification days for each of the 76 stations), for 72-h ahead forecast, they are based on 329764 verification cases (4339 verification days for each station).

\renewcommand{\arraystretch}{1.2}
\setlength{\tabcolsep}{1.2mm}
\begin{table}[!ht]\centering
\begin{tabular}{lcccc}
\toprule[0.09 em]
~ & CRPS & DSS & RMV & $\textup{Var}$(PIT)\\
\midrule
EMOS 48-h & 1.0101 & 2.4147 & 1.6156 & 0.0979 \\
AR-EMOS 48-h & 0.9897 & 2.1749 & 1.7263 & 0.0872 \\
\midrule
EMOS 72-h & 1.1244 & 2.6353 & 1.7831 & 0.099 \\
AR-EMOS 72-h & 1.0949 & 2.3548 & 1.9591 & 0.086 \\
\bottomrule[0.09 em]
\end{tabular}
\caption{Verification metrics of EMOS(ENS) and (heteroscedastic) AR-EMOS(ENS)  aggregated over $4340 \times 76$ (48-h) and $4339 \times 76$ (72-h) verification cases (dates and stations) \label{tab:verifAll48h72h}}
\end{table}

For both forecast horizons it is clearly visible, that AR-EMOS improves on EMOS in terms of CRPS and DSS, with the improvement being even more pronounced for 72-h ahead forecasts. Compared to the results on 24-h ahead forecasts, the improvement of AR-EMOS over EMOS becomes clearer the larger the forecast horizon.
For both considered horizons in Table \ref{tab:verifAll48h72h}, the EMOS predictive distribution is a bit sharper than its AR-EMOS counterpart, however, in each case the PIT variance of EMOS indicates underdispersion to a larger extent than the PIT variance of AR-EMOS.

\begin{figure}
\centering
\includegraphics[width=115mm, height=92mm]{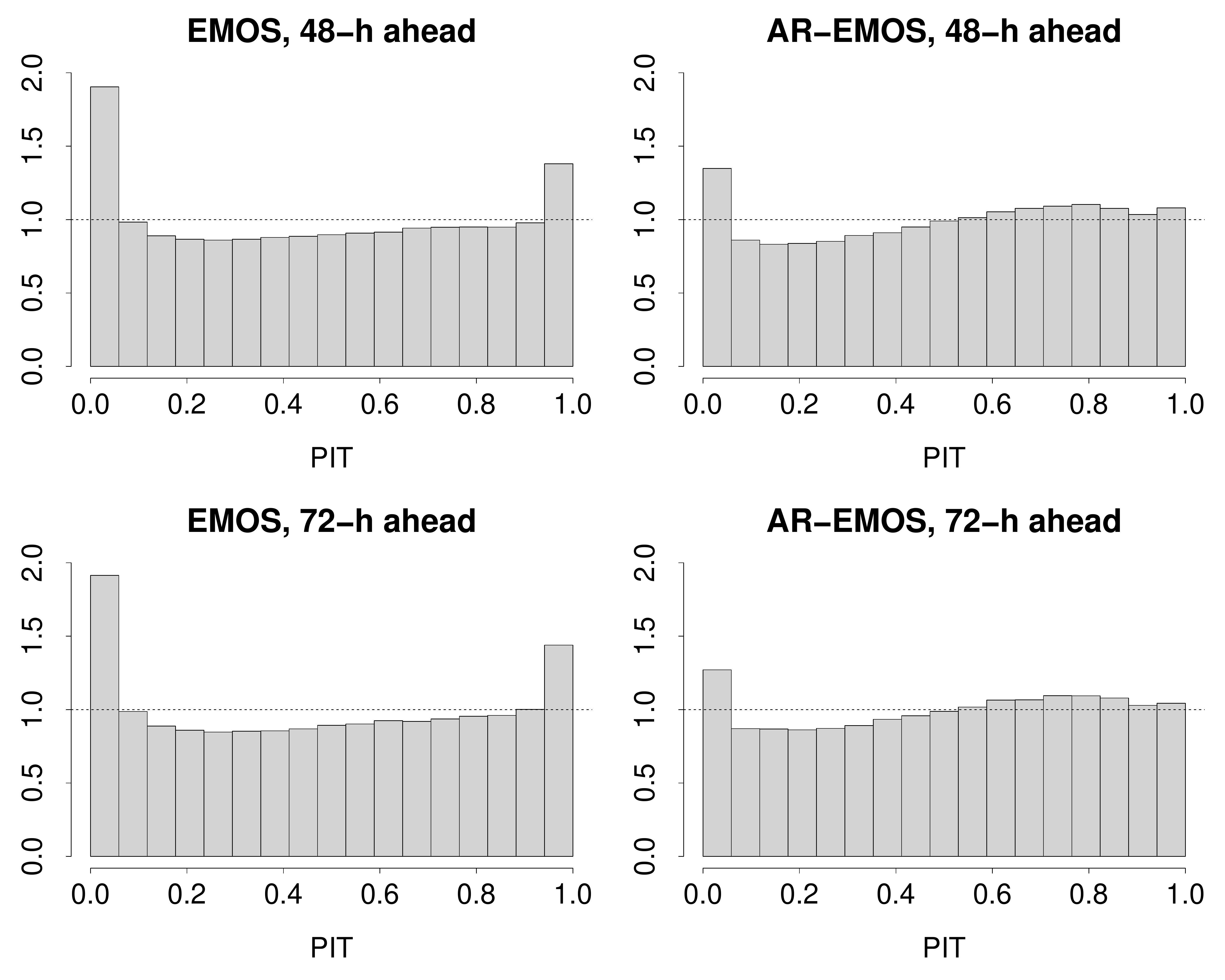}\caption{PIT histograms of EMOS and (heteroscedastic) AR-EMOS for 48-h ahead (top row), and for 72-h ahead (bottom row), aggregated over 76 stations, each with 4340 (48-h) and  4339 (72-h) verification dates \label{fig:PITAll48h72h}}
\end{figure}

Figure \ref{fig:PITAll48h72h} presents the respective PIT histograms of EMOS and AR-EMOS, where the top panel refers to 48-h ahead, and the bottom panel to 72-h ahead forecasts.

To investigate whether the improvement of AR-EMOS over EMOS is indeed a significant one, again the Diebold-Mariano test is performed at each station for the same one-sided alternative as in the previous paragraphs. The lower panel of Figure \ref{fig:DMtest} displays the resulting p-values for 48-h (left panel) and 72-h (right panel). For both forecast horizons the improvement in predictive performance of AR-EMOS over EMOS in terms of CRPS is highly significant for most of the stations: For 48-h ahead forecasts the p-value is $\leq 0.1$ at 64 stations (and $\leq 0.05$ still at 60 stations), for 72-h ahead forecasts even at 70 stations (and $\leq 0.05$ at 67 stations). In both cases, there is only one station with p-value greater than 0.9. Furthermore, the proportion of stations with an insignificant p-value is extremely small for both forecast horizons.

Although for 24-h ahead forecasts there is a larger portion of stations where AR-EMOS performs significantly better than EMOS, there is also a non-negligible number of stations where neither EMOS was significantly better than AR-EMOS nor the other way around. When moving to higher forecast horizons, the number of stations where AR-EMOS is significantly superior to EMOS increases heavily, and at more and more stations the level of significance even gets smaller. This indicates that the performance of AR-EMOS seems increases in comparison to EMOS for higher the forecast horizon.

\section{Conclusion}\label{sec:concl}

This follow-up work presents some new features and extensions of the AR-EMOS model introduced by \citet{moller2016probabilistic}, and is accompanied by an implementation of the method within an \texttt{R} package called \texttt{ensAR} \citep{package-ensAR}. The original model for the predictive variance is extended to incorporate the ensemble spread, yielding a heteroscedastic model implicitly accounting for the spread-error correlation, in slightly different way than the EMOS model. The (heteroscedastic) AR-EMOS model allows to fit a predictive distribution to a single ensemble member, as the longitudinal part of the (extended) predictive variance can still be computed for a single ensemble forecast, which is an advantage over standard postprocessing approaches such as EMOS.

Additionally, incorporation of a high-resolution forecast is investigated. The conducted case study indicates that this forecast improves predictive performance to a large extent. To incorporate the high-resolution forecast, an AR-EMOS group model is defined, which follows a somewhat different approach than the EMOS group model.
In the case study, only a simple heuristic form of the AR-EMOS group model is considered, which already yields excellent results. However, extensions to a more general and data driven form are relatively straight forward and subject to future research.

Finally, a feature of the AR-EMOS model not discussed in the original work is presented. The model allows to fit predictive distributions based on ensemble forecasts with arbitrary forecast horizons. In the original work a case study based only on 24-h ahead forecasts is presented. However, the AR-EMOS model can postprocess ensemble forecasts with arbitrary forecast horizons. For forecast horizons smaller or equal to 24-h ahead the model can be directly employed without any additional modifications. For horizons larger than 48-h ahead, the model can be applied by adding only one small preprocessing step, namely predicting the days between the last validation date of the forecast and the verification date with the AR model, also used to set up the AR-EMOS method itself.

The conducted case study indicates that for forecast horizons beyond 24-h ahead (with 48-h and 72-h ahead  considered as examples) the AR-EMOS performs particularly well, and improves significantly over EMOS. Therefore, the autoregressive postprocessing approach shows potential for extremely accurate prediction at higher forecast horizons.

\section*{Acknowledgements}

The authors acknowledge support of the research by Deutsche Forschungsgemeinschaft (DFG) Grant No. MO 3394/1-1, and by the Hungarian National Research, Development and Innovation Office under Grant No. NN125679.
Furthermore, we are grateful to the European Centre for Medium-Range Weather Forecasts (ECMWF) and the German Weather Service (DWD)
for providing forecast and observation data, respectively.

\end{document}